\documentclass[aps,twocolumn,amsfonts,amssymb,prb,superscriptaddress,10pt]{revtex4-2}

\usepackage{graphicx}
\usepackage{color}
\usepackage{xcolor}
\usepackage{amsfonts}
\usepackage{amsmath}
\usepackage[none]{hyphenat}
\usepackage{subfigure}
\usepackage[dvipsnames]{xcolor}
\usepackage[colorlinks=true,citecolor=blue]{hyperref}
\hypersetup{colorlinks=true,citecolor=blue,linkcolor=blue,urlcolor=blue}
\usepackage{outlines}

\allowdisplaybreaks 

\begin{document}

\title{Tomographic collective modes in a magnetic field} 

\author{Jeff Maki}
\email{jeffrey.maki@uni-konstanz.de}
\affiliation{Department of Physics, University of Konstanz, 78464, Konstanz, Germany}

\author{Johannes Hofmann}
\affiliation{Department of Physics, Gothenburg University, 41296 Gothenburg, Sweden}
\affiliation{Nordita, Stockholm University and KTH Royal Institute of Technology, 10691 Stockholm, Sweden}

\date{\today}

\begin{abstract}
Two-dimensional Fermi liquids at low temperatures have been theoretically established to exhibit an odd-even effect in the collective quasiparticle relaxation rates where even-parity deformations of the Fermi surface decay at a much faster rate than odd-parity ones. A predicted consequence of this effect is a new tomographic transport regime that mixes hydrodynamic and collisionless transport. In the presence of a magnetic field, however, the tomographic regime is expected to evolve towards conventional transport regimes as soon as the cyclotron radius becomes smaller than the dominant odd-parity mean free path. In this work, we examine this transition from the point of view of collective modes, using a numerically exact solution of the linearized Boltzmann equation within a generalized relaxation time approximation for the odd-parity and even-parity modes. In the absence of a magnetic field, the transverse conductivity exhibits two diffusive tomographic collective modes, and we find that at a critical magnetic field one of these two tomographic modes disappears. Which tomographic mode persists depends on the Landau parameters, and becomes increasingly dominated by hydrodynamic modes at high fields. We corroborate our analysis using a variational approach for the Fermi surface deformation that captures the angular structure of the deformation and the critical magnetic field strength. The collective modes discussed here can in principle be observed by examining the damping of longitudinal and transverse current responses in finite magnetic fields.
\end{abstract}

\maketitle

\section{Introduction}

The advent of ultra-clean materials has led to the observation of a new hydrodynamic transport regime in which momentum-conserving interactions between electrons dominate, leading to the formation of electronic fluids obeying the laws of hydrodynamics~\cite{Gurzhi1963, Gurzhi1968,deJong1995, Fritz2024,baker24,hui25}. Such electron-hydrodynamic materials are an exciting new paradigm in the theory of transport, in contrast to conventional transport that is predominately governed by momentum-relaxing scattering from impurities or phonons. 
The presence of prominent electron-electron interactions leads to viscous flow and vorticity \cite{Gurzhi1963,Gurzhi1968,deJong1995,Gurzhi1995,gurzhi_electron-electron_1995, Tomadin2014,Levitov2016, Scaffidi2017, Lucas2018, Cohen2018, Moessner2018, Alekseev2018,Hasdeo2021,Fritz2024, baker24,hui25}, ultraballistic transport \cite{Stern2022}, shear sound \cite{Semenyakin2018,khoo_shear_2019, Khoo2020,Valentinis2021,Valentinis2021b}, and magneto-hydrodynamic sound modes~\cite{Alekseev2019,Alekseev2020,kapralov_ballistic--hydrodynamic_2022,afanasiev_shear_2023}.  These findings are supported by experimental studies in ultraclean two-dimensional materials exploring viscous transport  properties \cite{deJong1995, Moll2016, Bandurin2016, Crossno2016, Nam2017, Gooth2018, Braem2018,Bandurin2018, Sulpizio2019, Kumar2022,AharonSteinberg2022, Jenkins2022, Krebs2023, Zeng2024, Palm2024, Moiseenko2025,kravtsov_viscous_2025,Geurs2025,Madhogaria2026}.

One important predicted fundamental manifestation of the mutual momentum-conserving interactions between electrons in two-dimensional (2D) materials is an odd-even effect in the quasiparticle relaxation: even-parity deformations of the quasiparticle distribution relax much faster than odd-parity ones at low temperatures \cite{Gurzhi1995,ledwith_hierarchy_2019, hofmann_anomalously_2023, nilsson_nonequilibrium_2024}. This odd-even effect is due to kinematic constraints on quasiparticle scattering in 2D Fermi liquids: as shown both analytically~\cite{Gurzhi1995, ledwith_hierarchy_2019} and by exact diagonalization of the linearized collision integral~\cite{hofmann_anomalously_2023,nilsson_nonequilibrium_2024}, even-parity deformations of the Fermi surface can relax via head-on collisions with a rate \mbox{$\gamma_e \sim k_B T^2/(\hbar T_F)$} (where \mbox{$T$} is the temperature, \mbox{$T_F$} is the Fermi temperature, \mbox{$\hbar$} is Planck's constant, and \mbox{$k_B$} is Boltzmann's constant) while below \mbox{$T\lesssim 0.15T_F$} selected odd-parity deformations (the number of which scales as \mbox{O($\sqrt{T_F/T}$)}) relax much more slowly, at an asymptotic rate \mbox{$\gamma_o \sim k_B T^4/(\hbar T_F^3)$}. A similar odd-even effect, albeit without the parametric change in the temperature dependence, has been shown to also exist in three-dimensional Fermi liquids \cite{Musser2026}. This hierarchy of relaxation rates leads to a ``tomographic'' transport limit \cite{ledwith_tomographic_2019}, in which odd-parity modes are collisionless while even-parity modes remain hydrodynamic.~Tomographic transport has been predicted to manifest in distinct transport signatures, including an anomalous wavenumber and temperature scaling of the transverse conductivity~\cite{ledwith_tomographic_2019,Kryhin2025,kryhin_two-dimensional_2023,rostami_magnetic-field_2024,Nazaryan2024}, additional diffusive collective modes \cite{Hofmann2022}, and enhanced boundary effects in finite-size geometries  \cite{ben-shachar_magnetotransport_2025,ben-shachar_tomographic_2025,Estradalvarez2025,Starkov2026}. 

A predicted key signature of tomographic transport that could discriminate the aforementioned effects from other transport mechanisms is their suppression in the presence of magnetic fields. For example, it has been shown that magnetic fields suppress tomographic scaling in the static transverse conductivity~\cite{rostami_magnetic-field_2024} as well as the tomographic boundary-layer structure in two-dimensional channel flow~\cite{ben-shachar_magnetotransport_2025,ben-shachar_tomographic_2025}. This is intuitive in general: when the cyclotron radius, \mbox{$r_c = v_F/\omega_c$} (where \mbox{$\omega_c\sim B$} is the cyclotron frequency and \mbox{$v_F$} is the Fermi velocity) is smaller than the mean-free-path of the dominant odd-parity mode \mbox{$\ell_o =v_F/\gamma_o$}, the relaxation of the odd-parity modes become suppressed and no signature of the odd-even effect in transport is expected. 

Detecting tomographic transport has so far been difficult owing to the presence of residual impurity effects and other momentum-relaxing processes that can obscure their transport signatures. Nevertheless, there has been recent progress to disentangle these momentum-relaxing processes in order to probe the anomalous scaling of the conductivity in the tomographic regime \cite{Zeng2024, Madhogaria2026}.
Another approach is to directly examine the angular momentum dependence of the relaxation rates. Here, the odd-even effect of the relaxation rates has been linked to the widths of cyclotron resonances of 2D Fermi gases \cite{Moiseenko2025}. It is therefore timely to further examine the odd-even effect and tomographic transport in the presence of a magnetic field.

In this article, we examine the transverse collective modes of a 2D Fermi liquid in the presence of a magnetic field and describe the crossover from the tomographic limit at small fields to conventional transport in the hydrodynamic and collisionless regimes at larger magnetic fields. We employ a numerical solution to the linearized Fermi liquid kinetic theory that captures not only the frequency of the tomographic collective modes but also the structure of the corresponding microscopic deformation of the Fermi surface. In the absence of a magnetic field, we also provide a variational \textit{Ansatz} for the Fermi surface deformation that accurately describes the collective mode structure. We find that at a critical magnetic field of order \mbox{$\omega_c/\gamma_o = {\it O}(1)$}, where \mbox{$\gamma_o$} is the dominant odd-parity relaxation rate of the collective mode, one of the two diffusive tomographic collective modes vanishes while the other persists as an odd-parity mode and eventually becomes increasingly dominated by hydrodynamic modes. Which tomographic mode vanishes depends on the Landau parameters, and we establish a critical Landau parameter at which both collective modes coalesce into a single diffusive mode. Intuitively, the critical field where the tomographic mode vanishes occurs when the cyclotron motion of quasiparticles in the dominant odd-parity modes becomes unaffected by odd-parity collisions. The dominant odd-parity mode of the Fermi surface of the collective motion depends on the variance of the angular momentum \mbox{$ \langle \sqrt{m^2} \rangle$}, leading to a critical magnetic field strength. When the odd-parity relaxation rate becomes similar to this angular momentum, the odd-parity mode becomes collisionless, and will not exhibit features of tomographic transport. This is explicitly captured in the variational \textit{Ansatz} for the Fermi surface deformation.

The article is organized as follows: In Sec.~\ref{sec:Kinetic_Theory}, we briefly revise the kinetic theory used to describe the transport properties and summarize the semi-analytical solution of the transport equations. In Sec.~\ref{sec:tom_modes}, we proceed to examine the collective mode spectrum in the presence of a magnetic field and discuss how the tomographic modes evolve with magnetic fields. We then employ a variational \textit{Ansatz} to examine the structure of the deformations of the Fermi surface of these modes and to estimate the critical magnetic field in Sec.~\ref{sec:variational}.
Our conclusions are presented in Sec.~\ref{sec:Discussions}.

\section{Kinetic Theory}
\label{sec:Kinetic_Theory}

\subsection{Fermi liquid kinetic equation}

We begin with a brief review of the solution to the Fermi liquid kinetic theory for 2D electron gases in a magnetic field~\cite{Pines2018, Baym1991,Lee1975,Simon1993,khoo_shear_2019, kapralov_ballistic--hydrodynamic_2022,Alekseev2018,Alekseev2019, afanasiev_shear_2023, Chiu1974}. The focus here is a concise derivation of the dynamical conductivity tensor that showcases the mixing of longitudinal and transverse modes in the response functions at finite magnetic fields. Our starting point is the Boltzmann equation for the local quasiparticle distribution function \mbox{$n_{\bf p}=n(t, {\bf r}, {\bf p})$} with momentum~\mbox{${\bf p}$} at position~\mbox{${\bf r}$} and time~\mbox{$t$},
\begin{equation}
\left[\partial_t + {\bf \dot{r}} \cdot \nabla_{\bf r} + {\bf \dot{p}} \cdot \nabla_{\bf p} \right] n_{\bf p} = I_{\rm coll}[n_{\bf p}] .
\label{eq:kinetic_equation}
\end{equation}
In Eq.~\eqref{eq:kinetic_equation}, \mbox{$\dot{\bf r}$} and \mbox{$\dot{\bf p}$} are the semiclassical equations of motion for a particle of charge \mbox{$-e$} in an electric field \mbox{${\bf E}({\bf r},t)$} and a magnetic field \mbox{${\bf B}({\bf r})$},
\begin{align}
    {\bf \dot{r}} &= \nabla_{\bf p} \epsilon_{\bf p}, \label{eq:rdot} \\[0.5ex]
    {\bf \dot{p}} &= -e{\bf E}({\bf r},t) - \nabla_{\bf r} \epsilon_{\bf p} - e {\bf \dot{r}} \times {\bf B}({\bf r}), \label{eq:pdot}
\end{align}
and $I_{\rm coll}[n_{\bf p}]$ is the collision integral, which accounts for quasiparticle relaxation by binary scattering. In writing Eq.~\eqref{eq:kinetic_equation}, we suppress the spin index, set \mbox{$\hbar = k_B = 1$}, and neglect effects due to a finite Berry curvature~\cite{Son2013,Niu1999}. The quasiparticle dispersion in Eq.~\eqref{eq:pdot} is given by
\begin{equation}
    \epsilon_{\bf p}= \frac{p^2}{2m^*} + \sum_{\bf p'}f_{{\bf p, p'}}\delta n_{\bf p'},
\end{equation}
where \mbox{$m^*$} is the renormalized effective quasiparticle mass.~The second term is a mean-field contribution to the quasiparticle energy, which is parametrized by the Landau function \mbox{$f_{\bf p, p'}$} with \mbox{$\delta n_{\bf p} = n_{\bf p} - n^0_{\bf p}$} denoting the deviation of the quasiparticle distribution from the global equilibrium Fermi-Dirac distribution \mbox{$n^0_{\bf p} = [e^{\beta (p^2/(2m^*)-\mu)}+1]^{-1}$} ($\mu$~is the chemical potential and  \mbox{$\beta = 1/T$} is the inverse temperature).

We consider the linear response of the system to a monochromatic electric field \mbox{${\bf E}({\bf r},t) = {\bf E}_0 e^{-i(\omega t+{\bf k} \cdot {\bf r})}$} (with \mbox{$\bf \hat{k} = \hat{x}$} in the following) in the presence of a homogeneous magnetic field in the out-of-plane \mbox{$z$}~direction, \mbox{${\bf B}({\bf r}) = B{\bf \hat{z}}$}. Linearizing Eq.~\eqref{eq:kinetic_equation} gives
\begin{align}
    \partial_t \delta n_{\bf p}  + &\left[{\bf v_p \cdot} \nabla_{\bf r} -\left({\bf v_p} \times e {\bf B}\right) \cdot \nabla_{\bf p}\right] \delta \bar{n}_{\bf p} - I_{\rm coll}[\delta \bar{n}_{\bf p}]\nonumber \\[0.5ex]
    &= - ({\bf v_p} \cdot e {\bf E}) \biggl(-\frac{\partial n^0_{\bf p}}{\partial \epsilon_{\bf p}}\biggr),
\label{eq:kinetic_eqn_linearized}
\end{align}
where \mbox{${\bf v_p} = \nabla_{\bf p}\epsilon_{\bf p}$} is the quasiparticle velocity and \mbox{$\delta \bar{n}_{\bf p} = \sum_{\bf p'} (\delta_{\bf p,p'} -\frac{\partial n^0_{\bf p}}{\partial \epsilon_{\bf p}}) f_{\bf p,p'}) \delta n_{\bf p'}$} is the deviation of the distribution function from local equilibrium.

Due to the rotational symmetry of the problem, we expand the distribution function and the quasiparticle interaction in terms of angular harmonics. Denoting the polar angle of \mbox{$\bf p$} with respect to the coordinate \mbox{$x$}~axis by~\mbox{$\theta_{\bf p}$} we write
\begin{align}
    \delta n_{\bf p} &= \biggl(-\frac{1}{\beta}\frac{\partial n_{\bf p}^0}{\partial \epsilon_{\bf p}}\biggr)h(\theta_{\bf p}) \label{eq:h}
\end{align}
with
\begin{align}
    h(\theta_{\bf p}) &= \sum_{m} h_me^{i m\theta_{\bf p}} , \label{eq:expansion_n} 
\end{align}
and  \mbox{$m \in \mathbb{Z}$}. Here, \mbox{$h(\theta_{\bf p})$} is a dimensionless function that describes a rigid Fermi surface deformation, where odd \mbox{$m$} describe an odd-parity deformation (i.e.\,~a deformation for which \mbox{$h(-\theta_{\bf p}) = -h(\theta_{\bf p})$}), and even \mbox{$m$} an even-parity deformation (a deformation for which \mbox{$h(-\theta_{\bf p}) = h(\theta_{\bf p})$}).  Likewise, the Landau function is also expanded in terms of angular harmonics
\begin{align}
    f_{\bf p,p'} &= \frac{1}{\nu}\sum_m F_m e^{i m (\theta_{\bf p}-\theta_{\bf p'})} ,
\end{align}
where \mbox{$\nu = m^*/\pi$} the density of states of a two-component electron gas. The renormalized mass depends on the \mbox{$p$}-wave Landau parameter as \mbox{$m^*/m = (1+F_1)$}. These relations imply \mbox{$\bar{h}_m = (1+F_m)h_m$}. Finally, for the linearized collision integral
\begin{align}
I_{\rm coll}[\delta \bar{n}_{\bf p}] &= - \biggl(-\frac{1}{\beta}\frac{\partial n^0_{\bf p}}{\partial \epsilon_{\bf p}}\biggr) L[\bar{h}(\theta_{\bf p})] ,
\end{align}
by rotational invariance each angular harmonic \mbox{$h_m$} has its own relaxation rate~\mbox{$\gamma_m$},
\begin{align}
L[\bar{h}(\theta_{\bf p})] &= \sum_m \gamma_m \bar{h}_m e^{i m \theta_{\bf p}} .
\label{eq:collision_integral_expansion} 
\end{align}
In a minimal model of the odd-even effect, the relaxation rates \mbox{$\gamma_m$} have the form
\begin{equation}
\gamma_m = \begin{cases}
0 & m=0 \\
0 & |m| = 1 \\
\gamma_e & \text{$m$ is even and $|m|\geq2$}\\
\gamma_o(m)  &\text{$m$ is odd and $|m|\geq3$} .
\end{cases}
\label{eq:relaxation_rates}
\end{equation}
The zero modes \mbox{$\gamma_{m=0}=0$} and \mbox{$\gamma_{m=\pm1}= 0$} are related to the conservation of particle number and momentum in two-body collisions, respectively. 

The simplest model that captures  the essential physics of the tomographic regime assumes that the odd-parity relaxation rate is independent of angular momentum~\cite{Hofmann2022}
\begin{equation}
\gamma_o(m) = \gamma_o 
\label{eq:constant_gammaodd}
\end{equation}
with \mbox{$\gamma_o \ll \gamma_e$}. 
A more refined model assumes angular dependent odd-parity relaxation rates of the form \cite{rostami_magnetic-field_2024,nilsson_nonequilibrium_2024,Kryhin2025}
\begin{equation}
    \gamma_{o}(m) = \frac{1}{\frac{1}{\gamma_e} + \frac{3^4}{\gamma_o m^4}},
    \label{eq:m_dependent_gammaodd}
\end{equation}
which states that, even for \mbox{$\gamma_o \ll \gamma_e$}, only odd-parity modes with small angular momenta exhibit the odd-even effect. For such small angular momenta, the odd-parity relaxation rates scale as \mbox{$\gamma_o(m)\sim \gamma_o (m/3)^4$} \cite{ledwith_hierarchy_2019, ledwith_tomographic_2019,Hofmann2022, Kryhin2025}. This scaling is terminated at large angular momentum for which the odd-parity and even-parity relaxation rates are nearly identical~\cite{nilsson_nonequilibrium_2024}. We will use both models in the remainder of this paper. Both give qualitatively similar results, and we use the simpler odd-even staggered model~\eqref{eq:constant_gammaodd} when our calculations can be compared to analytical results for this model.

Putting everything together gives the final result for the Boltzmann kinetic equation
\begin{align}
-i[\omega &+i \gamma_m + m(1+F_m)\omega_c]h_m\nonumber \\[0.5ex]
&\qquad + \frac{iv_Fk}{2} [(1+F_{m+1})h_{m+1}+(1+F_{m-1})h_{m-1}] \nonumber \\
&\quad = \frac{v_F e E_0}{2}\left(
    e^{i\theta_E}\delta _{m,1} + e^{-i \theta_E}\delta_{m,-1}\right),
\label{eq:Boltzmann_m_E}
\end{align}
where \mbox{$\omega_c = e B/m^*$} is the cyclotron frequency in terms of the effective quasiparticle mass and \mbox{$\theta_E$} is the angle of the electric field with the reference $x$~axis. For longitudinal electric fields \mbox{$\theta_E=0$} (i.e.\ \mbox{${\bf E} \parallel {\bf k}$}), while for transverse fields \mbox{$\theta_E=\pi/2$} (i.e.\ \mbox{${\bf E} \perp {\bf k}$}). For explicit calculations, we will include the \mbox{$p$}-wave Landau parameter \mbox{$F_1$}, which determines the properties of the transverse collective modes. 

\subsection{Semi-analytical solution of the kinetic equation}

An efficient way to analyze Eq.~\eqref{eq:Boltzmann_m_E} is to note that for \mbox{$|m|\geq 2$}, \mbox{$h_m$} satisfies a recursive relation
\begin{align}
    &-i\bigl[\omega + i \gamma_m + m(1+F_m)\omega_c\bigr] h_m \nonumber \\[2ex]
    & +\frac{iv_Fk}{2}\bigl[(1+F_{m+1})h_{m+1} + (1 + F_{m-1}) h_{m-1}\bigr] = 0 ,
    \label{eq:higher_m}
\end{align}
without reference to the external drive. This implies that \mbox{$h_{|m|\geq2}$} is a fixed function of the external frequency $\omega$ and momentum $k$, which can be used to determine \mbox{$h_{0,\pm1}$}, as shown in Appendix~\ref{app:x_2}. Defining 
\begin{equation}
    x_{\pm m} = \frac{h_{\pm m}}{h_{\pm (|m|-1)}} , \label{eq:continuedfraction}
\end{equation}
the linear set of equations to describe the \mbox{$m=0,\pm 1$} subspace then reads
\begin{widetext}
\begin{align}
    \begin{bmatrix}
        -i (\omega + (1+F_1)\omega_c) + \frac{iv_Fk}{2}x_2 & \frac{iv_Fk}{2}(1+F_0) & 0 \\[0.5ex]
        \frac{iv_Fk}{2}(1+F_1) & -i \omega & \frac{iv_Fk}{2} (1+F_1) \\[0.5ex]
        0 & \frac{iv_Fk}{2}(1+F_0) & -i (\omega -(1+F_1)\omega_c) + \frac{iv_Fk}{2}x_{-2}
    \end{bmatrix} & \begin{bmatrix}
        h_1 \\
        h_0 \\
        h_{-1} 
    \end{bmatrix} 
    = \frac{\beta v_FeE_0}{2}\begin{bmatrix}
         e^{i\theta_E} \\
        0 \\
         e^{-i \theta_E}
    \end{bmatrix} , \label{eq:reduced_kinetic_theory}
\end{align}
\end{widetext}
where we set \mbox{$F_{m\geq2}=0$} for the remainder of this work.
Inverting Eq.~\eqref{eq:reduced_kinetic_theory} gives an analytical expression for $h_{m = 0, \pm1}$ in terms of $x_{\pm2}$,  which is used to evaluate the charge current
\begin{equation}
    {\bf j} = 2 (-e) \int \frac{d^2p}{(2\pi)^2} {\bf v_p} \delta \bar{n}_{\bf p} = \sigma {\bf E}_0,
\end{equation}
where $\sigma$ is the conductivity tensor,
\begin{align}
\frac{k^2 \sigma(\omega, {\bf k})}{\nu e^2} = -2i\frac{(1+F_1)(v_Fk)^2 \omega}{s_ls_t + s_b^2} \begin{bmatrix}
s_t & s_b \\
-s_b & s_l
\end{bmatrix} .
\label{eq:sigma}
\end{align}
In Eq.~\eqref{eq:sigma}, we define the following functions
\begin{align}
s_l &= (1+F_0)(1+F_1)(v_Fk)^2 +  v_Fk \omega x_+ - 2\omega^2, \label{eq:dl} \\[0.5ex]
s_t &=v_Fk \omega x_+ -2\omega^2, \label{eq:dt}\\[0.5ex]
s_b &= i \omega (-2(1+F_1)\omega_c +v_Fkx_-) , \label{eq:db}
\end{align}
as well as
\begin{align}
    x_{\pm} = x_2 \pm x_{-2}. \label{eq:x2def}
\end{align}
Note that \mbox{$x_-$} vanishes as a linear function of \mbox{$\omega_c$} in the limit of small magnetic fields, while \mbox{$x_+$} approaches a fixed function of momentum \mbox{$k$} and frequency \mbox{$\omega$}.
Explicit expressions and additional limits of \mbox{$x_\pm$} are discussed in App.~\ref{app:x_2}. 
With our convention \mbox{$\hat{\bf k} = \hat{\bf x}$}, the longitudinal and  transverse conductivities are given by
\begin{align}
    \sigma_L(\omega,{\bf k}) &= \sigma_{xx}(\omega,{\bf k}) , \\[0.5ex]
    \sigma_T(\omega,{\bf k}) &= \sigma_{yy}(\omega,{\bf k}),
    \label{eq:conductivities}
\end{align}
while the Hall conductivity is
\begin{equation}
    \sigma_H(\omega,{\bf k}) = \sigma_{xy}(\omega,{\bf k}).
\end{equation}

In this approach, all information of the higher angular momentum modes is contained in the functions~\mbox{$x_{\pm}$} [Eq.~\eqref{eq:x2def}]. We remark that this is different compared to other treatments of the kinetic equation~\eqref{eq:kinetic_eqn_linearized} that work in a reference frame that is rotating at the cyclotron frequency~\cite{Lee1975,Simon1993, Moiseenko2025}. Both approaches are in principle equivalent. The approach employed here is more convenient for numerically evaluating the collective mode spectrum, and we will use it below.

\subsection{structure of the conductivity tensor}
\label{subsec:struct}

\begin{figure}
    \centering
    \includegraphics{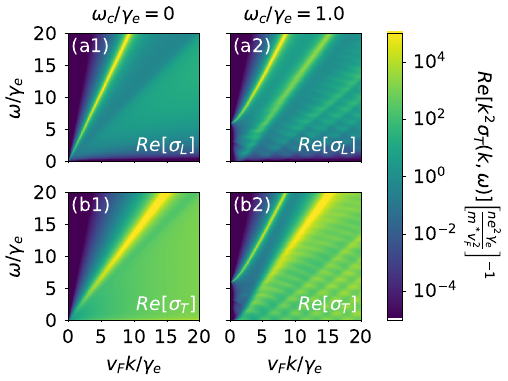}
    \caption{Real part of the longitudinal conductivity ${\rm Re}[\sigma_L]$ [top row, panels (a)] and the transverse conductivity ${\rm Re}[\sigma_T]$ [bottom row, panels (b)] calculated with the odd-even staggered relaxation rate model in [Eq.~\eqref{eq:constant_gammaodd}]. The left column [panels (1)] shows the response in the absence of magnetic field, and the right column [panels (2)] in the presence of a finite magnetic field with \mbox{$\omega_c/\gamma_e = 1$}. We set \mbox{$F_1 =5$} and \mbox{$\gamma_o/\gamma_e=10^{-2}$}, and the  unit of the conductivity is \mbox{$ne^2\gamma_e/(m^*v_F^2)$} with \mbox{$n = k_F^2/\pi$} the density.}
    \label{fig:1}
\end{figure}


To illustrate the structure of the current response, Fig.~\ref{fig:1} shows a numerical solution for the (a) longitudinal and (b) transverse parts of the conductivity tensor, ${\rm Re}[\sigma_L]$ and ${\rm Re}[\sigma_T]$, respectively, both in (1) the absence and (2) the presence of a magnetic field. The figure is arranged such that the top (bottom) row shows ${\rm Re}[\sigma_L]$ (${\rm Re}[\sigma_T]$), and the left (right) panels the response in the absence (in the presence) of a magnetic field. Note that in the limit of zero magnetic field \mbox{$\omega_c\to0$}, since \mbox{$x_- \sim \omega_c$,} the off-diagonal elements of the conductivity tensor vanish, \mbox{$s_b=0$}, as required by time-reversal invariance. Only the longitudinal and transverse response [shown in Figs.~\ref{fig:1}(a1) and~\ref{fig:1}(b1)] remain finite with (from Eq.~\eqref{eq:sigma})
\begin{align}
\frac{k^2 \sigma(\omega,{\bf k})}{\nu e^2} = -2i(1+F_1)(v_Fk)^2 \omega\begin{bmatrix}
s_l^{-1} & 0 \\
0 & s_t^{-1}
\end{bmatrix}. 
\label{eq:sigma_noB}
\end{align}
Note that the longitudinal response in the absence of a magnetic field is governed by the function \mbox{$s_l$} [cf.~Eq.~\eqref{eq:dl}] while for the transverse response by \mbox{$s_t$} [cf.~Eq.~\eqref{eq:dt}]. This implies that the collective modes of the longitudinal and transverse differ from one another due to the different structure of the functions \mbox{$s_l$} and \mbox{$s_t$}. This separation is directly apparent in the left column of Fig.~\ref{fig:1}:  the longitudinal response [Fig.~\ref{fig:1}(a1)] shows a clear sound mode with linear dispersion that is separated from the particle-hole continuum at \mbox{$\omega = v_F k$}~\cite{Pines2018,Baym1991}.
This mode exists both in the hydrodynamic (\mbox{$v_F k/\gamma_e\ll 1$}) and collisionless (\mbox{$v_Fk/\gamma_e \gg 1$}) limits.
The transverse response [Fig.~\ref{fig:1}(b1)], by contrast, has a well defined shear sound mode only in the collisionless limit, with a different velocity from the longitudinal sound mode.  In the hydrodynamic limit the transverse mode becomes indistinguishable from the particle-hole continuum~\cite{khoo_shear_2019}, and is purely diffusive.

In the presence of a magnetic field [Fig.~\ref{fig:1}, panels 2],
the previously separated longitudinal [Fig.~\ref{fig:1}(a2)] and transverse response [Fig.~\ref{fig:1}(b2)] become mixed at finite momenta and frequencies, as implied by Eq.~\eqref{eq:sigma}, with now two well-defined collective mode features in each response function in the collisionless regime (i.e. for~\mbox{$v_Fk/\gamma_e\gg 1$}). In addition, the particle-hole continuum in both the longitudinal and transverse response stratifies into a series of cyclotron resonances. 

In addition, at zero frequency and finite momentum, the conductivity tensor is diagonal even in the presence of a magnetic field [i.e.\,~\mbox{$s_b=0$}, cf.\,~Eq.~\eqref{eq:db}], and the longitudinal and transverse components are decoupled. One signature of this is that the zero-frequency response at finite wavenumber in Fig.~\ref{fig:1}1 and 2 exhibits a similar reduction of the particle-hole continuum in the longitudinal response, while the transverse response remains finite. In the opposite limit of zero momentum and finite frequency, there is a single finite response at the first renormalized cyclotron resonance~\mbox{$\omega= (1+F_1)\omega_c$}. Note that, although it appears as if this frequency depends on the quasiparticle interactions, the cyclotron frequency itself depends on the renormalized quasi-particle mass~\mbox{$m^*$}. The dependence on \mbox{$F_1$}  thus cancels, as expected, leading to a response at the bare cyclotron frequency~\cite{Kohn1961}.
As the frequency approaches zero, \mbox{$\omega/\gamma_e \to 0$}, at zero momentum, the transverse and longitudinal conductivities become vanishingly small and lead to the classical Hall effect \cite{Lee1975}. In this limit the conductivity tensor takes the form \cite{Giuliani2005}
\begin{equation}
    \lim_{\omega \to 0} \lim_{k \to 0} \sigma = \frac{ne}{B} \begin{bmatrix}
        0 & 1 \\
        -1 & 0
    \end{bmatrix}.
    \label{eq:classical_hall}
\end{equation}

\begin{figure}
    \centering
    \includegraphics{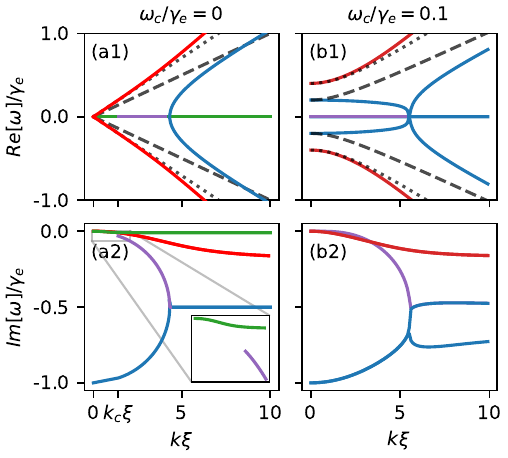}
    \caption{Collective mode spectrum for \mbox{$F_1 = 5$} and \mbox{$\gamma_o/\gamma_e = 10^{-2}$} in the (a) absence and (b) presence of a magnetic field of strength \mbox{$\omega_c/\gamma_e = 10^{-1}$} (left and right columns). The top (bottom) row shows the (1)~real mbox{$Re[\omega]$} and (2)~and imaginary \mbox{$Im[\omega]$} parts. Red lines describe the longitudinal sound mode. The blue lines denote transverse sound modes, which become propagating at finite \mbox{$k\xi$} \cite{khoo_shear_2019}. The  green and purple lines denote upper and lower branches of the tomographic collective modes Eq.~\eqref{eq:frequencies_exact}. The inset in (a,ii) highlights the branch cut between these two modes. The black dashed and dotted lines show the analytical predictions for (a) the longitudinal and transverse sound modes and (b) the magnetoplasmon and magnetosound modes, Eqs.~\eqref{eq:mp}--\eqref{eq:ms}.}
    \label{fig:2}
\end{figure}

\section{Collective Modes of the Conductivity Tensor}
\label{sec:tom_modes}

We now turn to a detailed discussion of the longitudinal and transverse collective mode spectrum.

\subsection{Analysis of the collective mode spectrum}

To examine the collective mode spectrum, we  evaluate the poles of the longitudinal or transverse conductivities:
\begin{equation}
    \sigma_{L,T}^{-1}({\bf k},\omega) = 0.
\end{equation}

The evolution of the real \mbox{$Re[\omega]$}(top row) and imaginary \mbox{$Im[\omega]$} (bottom row) parts of the pole structure is shown in Fig.~\ref{fig:2} in the [Fig.~\ref{fig:2}(a)] absence and [Fig.~\ref{fig:2}(b)] presence of a magnetic field (left and right columns) as a function of the dimensionless quantity~\mbox{$k\xi$}, where \mbox{$\xi$} is the characteristic length scale 
\begin{equation}
    \xi = \frac{v_F}{\sqrt{\gamma_e\gamma_o}}.
\end{equation}
This scale parametrizes the onset of the tomographic regime at \mbox{$k\xi \sim 1$}~\cite{ledwith_tomographic_2019,Hofmann2022}, whereas for \mbox{$k\xi \ll 1$} the system is in the hydrodynamic regime and for \mbox{$k\xi \gg 1$} in the collisionless regime. 

\subsubsection{Collective-mode structure at zero field}

In the hydrodynamic limit \mbox{$k\xi\ll 1$}, the real part of the collective mode spectrum [Fig.~\ref{fig:2}(a)] exhibits a linearly dispersive and weakly damped longitudinal sound mode (red line in Fig.~\ref{fig:2}), which exists for all momenta and evolves into the collisionless zero sound mode at large momenta (\mbox{$k\xi \gg 1$}). For comparison, we indicate the hydrodynamic first-sound dispersion \mbox{$\omega_{L} \approx \pm \sqrt{(1+F_1)/2} v_Fk -i(v_Fk)^2/(8\gamma_e)$} by a black dotted line in the figure. 
The transverse collective mode is indicated by the blue line. It is purely diffusive in the hydrodynamic regime, with a damping rate that starts at \mbox{$-\gamma_e$}, and it becomes propagating in the collisionlesss limit. The green line represents another diffusive transverse sound mode with a frequency \mbox{$\omega = -i(1+F_1)(v_Fk)^2/(4\gamma_e)$}, which becomes vanishingly small in the hydrodynamic limit.  Its continuation at finite momentum has been dubbed the upper branch of the tomographic mode~\cite{Hofmann2022}.

In the tomographic regime \mbox{$k\xi \simeq 1$}, two diffusive poles appear, which are indicated by the green and purple lines in Fig.~\ref{fig:2}(a2). These are the two diffusive tomographic collective modes~\cite{Hofmann2022}. For the staggered odd-even relaxation rates in Eq.~\eqref{eq:constant_gammaodd}, our results agree with an analytical expression for the dispersion of these diffusive poles~\cite{Hofmann2022}
\begin{align}
    \omega_{tom,\pm}(k) &= -i \gamma_o \frac{1+F_1}{8F_1} \biggl[ 4 + \left(k\xi\right)^2 (1+F_1) \nonumber \\
    & \mp (k\xi)^2 \sqrt{\left(1+F_1 +\frac{4}{(k\xi)^2}\right)^2- \frac{16F_1}{(k\xi)^2}}\biggr].
    \label{eq:frequencies_exact}
\end{align}
As discussed above, only the upper branch (green line) exists at small momentum.
When \mbox{$k\xi$} is increased beyond a threshold value \mbox{$k_c\xi = \sqrt{2/(F_1-1)} $} (\mbox{$k_c\xi = 1.4$} for the particular values in Fig.~\ref{fig:2}, as indicated in the figure axis) the lower branch mode emerges, which is denoted by the purple line. Conversely, for fixed momentum, the lower branch will appear above a critical Landau parameter~\cite{Hofmann2022}
\begin{equation}
    F_c = 1 + \frac{2}{(k\xi)^2}.
\end{equation} 
The two branches are separated by a branch cut along the imaginary frequency axis of length \mbox{$\omega \in [-i\gamma_o, -i\gamma_o(1+(k\xi)^2)]$}. This cut can be observed in Fig.~\ref{fig:2}(a2) (highlighted in the inset) as a small gap between the green and purple lines near \mbox{$k_c\xi$}.

Finally, in the collisionless limit \mbox{$k\xi \gg 1$}, the upper branch of the tomographic mode approaches zero damping. The lower branch, by contrast, terminates at a finite value of \mbox{$k\xi$}, at which point it coalesces with the hydrodynamic diffusive mode (blue line) and turns into a dispersive shear sound mode with \mbox{$\omega_T \approx \pm \sqrt{(1+F_1)/4}v_Fk-i \gamma_e/2$} (black dashed line)~\cite{khoo_shear_2019, afanasiev_shear_2023}.

\subsubsection{Collective-mode structure at finite magnetic fields}

At finite magnetic fields, Fig.~\ref{fig:2}(b), the longitudinal and transverse sound modes are replaced by the magnetoplasmon (mp) and magnetosound (ms) modes~\cite{afanasiev_shear_2023, khoo_shear_2019,Alekseev2018,Alekseev2019, kapralov_ballistic--hydrodynamic_2022}. In the absence of the odd-even effect, their dispersions are predicted as
\begin{align}
    Re[\omega_{\rm mp}] &\approx \sqrt{(1+F_1)^2 \omega_c^2 +(1+F_1)\frac{(v_Fk)^2}{2}}, \label{eq:mp} \\
    Re[\omega_{\rm ms}] &\approx \sqrt{4 \omega_c^2 +(1+F_1)\frac{(v_Fk)^2}{4}}. \label{eq:ms} 
\end{align}
These two dispersions are shown as black dotted lines and black dashed lines in Fig.~\ref{fig:2}(b1). Compared to the numerical results (red lines for magnetoplasmon and blue lines for the magnetosound), it is apparent that these expressions capture the hydrodynamic and the collisionless regimes, but not the intermediate tomographic regime. In the hydrodynamic limit \mbox{$k\xi \ll 1$}, these collective modes start at the first and second renormalized cyclotron resonance, respectively. In the collisionless regime \mbox{$k\xi \gg 1$}, the magnetoplasmon and magnetosound modes asymptotically connect to the longitudinal and transverse sound modes discussed previously.

The damping rate of the collective modes in the presence of a magnetic field is shown in Fig.~\ref{fig:2}(b2). For this choice of magnetic fields, we see no qualitative change to the damping rate of the longitudinal modes~\cite{Hofmann2022}. In terms of the transverse modes, we still observe an exceptional point (near \mbox{$k\xi\simeq 5$}) at which these modes become propagating. Notably at small momenta there is only one tomographic mode, the lower branch [purple line in Fig.~\ref{fig:2}(b2)]. In the hydrodynamic limit \mbox{$k\xi \to 0$}, this mode continuously approaches zero damping without any appearance of the branch cut, in contrast to the zero-field case in which this modes disappears at finite momentum [cf.~Fig.~\ref{fig:2}(a2)]. This conforms with the expectation that the magnetic field suppresses tomographic transport signatures. In the following, we proceed to analyze in more detail the disappearance of the tomographic regime with \mbox{$k\xi \sim 1$} and the evolution of the collective mode spectrum in a magnetic field.

\begin{figure}
    \centering
    \includegraphics{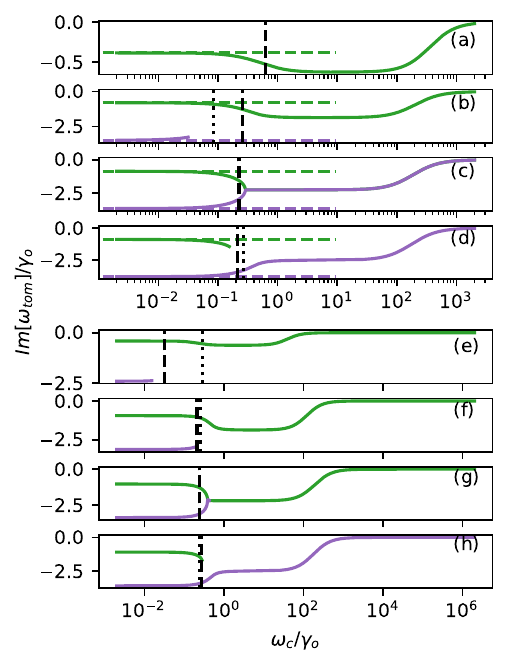}
    \caption{Spectrum of the two tomographic diffusive collective modes of the transverse conductivity, \mbox{$\omega_{tom}$}, as a function of the magnetic field strength \mbox{$\omega_c/\gamma_o$}. (a)--(d) are obtained assuming a constant odd-parity relaxation rate [Eq.~\eqref{eq:constant_gammaodd}], for (a)~\mbox{$F_1=0.0$}, (b)~\mbox{$F_1=2.0$}, (c)~\mbox{$F_1=2.6$}, and (d)~\mbox{$F_1=3.0$}. We set \mbox{$v_Fk/\gamma_e = 5\cdot10^{-3}$} and \mbox{$\gamma_o/\gamma_e = 10^{-5}$}, corresponding to \mbox{$k\xi = 1.58$}. The green and purple lines denote the upper and lower branches, which at zero magnetic field coincide with the predictions of Eq.~\eqref{eq:frequencies_exact} [horizontal dashed lines]. The vertical black dashed (dotted) lines are estimates for the critical magnetic field obtained from Eq.~\eqref{eq:critical_cond}. (e)--(h) similar to (a)--(d) but using the \mbox{$m$}-dependent relaxation rates in Eq.~\eqref{eq:m_dependent_gammaodd}. We use \mbox{$v_Fk/\gamma_e=10^{-2}$} and \mbox{$\gamma_o/\gamma_e\approx 0.8 \times 10^{-5}$} with (e)~\mbox{$F_1=0$}, (f)~\mbox{$F_1=2.0$}, (g)~\mbox{$F_1=2.5$}, and (h)~\mbox{$F_1=3.0$}.}
    \label{fig:3}
\end{figure}

\begin{figure}
    \centering
    \includegraphics{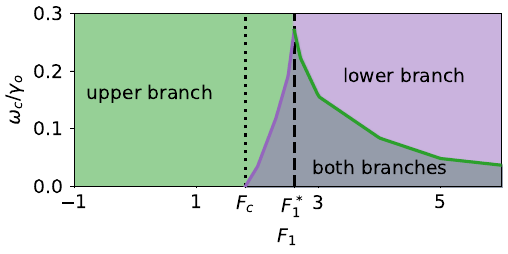}
    \caption{Stability diagram indicating the presence of tomographic modes as a function of~\mbox{$F_1$} and the magnetic field for the linearized collision integral with constant odd-parity relaxation rates [Eq.~\eqref{eq:constant_gammaodd}]. For the green (purple) shaded region there exists an upper (lower) branch tomographic mode, with an overlap region in which both modes are present. The solid green and purple lines depict the critical values of \mbox{$\omega_c/\gamma_o$}. The black dotted and dashed lines denote the point~\mbox{$F_c$} at which the lower branch emerges, and the point~\mbox{$F_1^*$} at which the instability point of the two collective modes meet. We choose the same parameters as in Fig.~\ref{fig:3} and set \mbox{$v_Fk/\gamma_e = 5\cdot10^{-3}$}  and \mbox{$\gamma_o/\gamma_e = 10^{-5}$} (\mbox{$k\xi \approx 1.58 $}). }
    \label{fig:4}
\end{figure}
 
In Fig.~\ref{fig:3}, we explore the evolution of the diffusive collective modes in the tomographic limit in the presence of a magnetic field for a staggered odd-parity relaxation rate Eq.~\eqref{eq:constant_gammaodd} for \mbox{$F_1 = 0,2.0,2.6,3.0$} in Figs.~\ref{fig:3}(a)--(d), while in Fig.~\ref{fig:3}(e)--(h) we employ the \mbox{$m$}-dependent odd-parity relaxation rate in Eq.~\eqref{eq:m_dependent_gammaodd} for \mbox{$F_1 = 0, 2.0,2.5,3.0$}. The solid green and purple lines denote the upper and lower branches, respectively, for a fixed momentum \mbox{$v_Fk/\gamma_e = 5 \cdot 10^{-3}$} or equivalently \mbox{$k\xi \approx 1.58$}.

For Figs.~\ref{fig:3}(a)--(d), the spectrum of the tomographic modes does not change appreciably for \mbox{$\omega_c/\gamma_o \lesssim 10^{-1}$}. The frequency obtained for the upper and lower branch in this regime agrees with the analytical results in Eq.~\eqref{eq:frequencies_exact}, which are marked by the green and purple dashed lines, respectively. Thus, for weak magnetic fields there are still two tomographic diffusive modes (provided, of course, that \mbox{$F_1>F_c$}). 

As is apparent in all panels, with increasing magnetic field there is a critical value at which one of the tomographic collective modes vanishes while the damping of the remaining mode changes substantially compared to the zero magnetic field limit. In particular, the frequency of the remaining mode either decreases for the upper branch while it increases for the lower branch. We find that the value of the Landau parameter \mbox{$F_1$} dictates whether the upper or lower branch vanishes at the critical magnetic field strength: for small values of \mbox{$F_1$} the lower branch disappears, while for large values of \mbox{$F_1$} it is the upper branch that disappears. This defines a second critical Landau parameter \mbox{$F_1^*$} at which the two branches coalesce at the critical magnetic field strength. This is shown in Fig.~\ref{fig:3}(c), where \mbox{$F_1^* = 2.6$} for our parameters. Further increasing the magnetic field past this point, the frequency of the stable mode remains unchanged over several orders of magnitude of \mbox{$\omega_c/\gamma_o$}, and then eventually transitions towards zero frequency. This final transition occurs well before~\mbox{$\omega_c \sim \gamma_e$}.

In Figs.~\ref{fig:3}(e)--(h), which corresponds to the \mbox{$m$}-dependent odd-parity relaxation rates in Eq.~\eqref{eq:m_dependent_gammaodd}, we observe no major qualitative difference compared to the results from a constant and \mbox{$m$}-dependent odd-parity relaxation rate, Eq.~\eqref{eq:constant_gammaodd}. The only change is the quantitative value of the critical Landau parameter, \mbox{$F_1^*$}. For the  parameters chosen in Figs.~\ref{fig:3}(e-h) the critical value of the Landau parameter is \mbox{$F_1^* \approx 2.5$}.

The stability diagram for the existence of the tomographic mode branches assuming a constant odd-parity relaxation rate is shown in Fig.~\ref{fig:4}. The green (purple) shaded regions are parameter regimes in which the upper (lower) branch of the tomographic collective mode exists. The solid lines denote the critical fields at which one mode terminates [cf.~Fig.~\ref{fig:3}]. At small values of \mbox{$F_1$} the lower branch disappears at a finite magnetic field strength. For \mbox{$F_1^* =2.6$}, the upper and lower branches meet, as seen in Fig.~\ref{fig:3}(c). Beyond this point, it is the upper branch that disappears at finite magnetic fields, while the lower branch persists. We find that the explicit value of \mbox{$F_1^*$} depends on the specific form of the relaxation rates, and the value of the momentum~\mbox{$k$}. However, this will only quantitatively change the transition point, and therefore we refrain from mapping out \mbox{$F_1^*(k)$}.

\begin{figure*}
    \centering
    \includegraphics{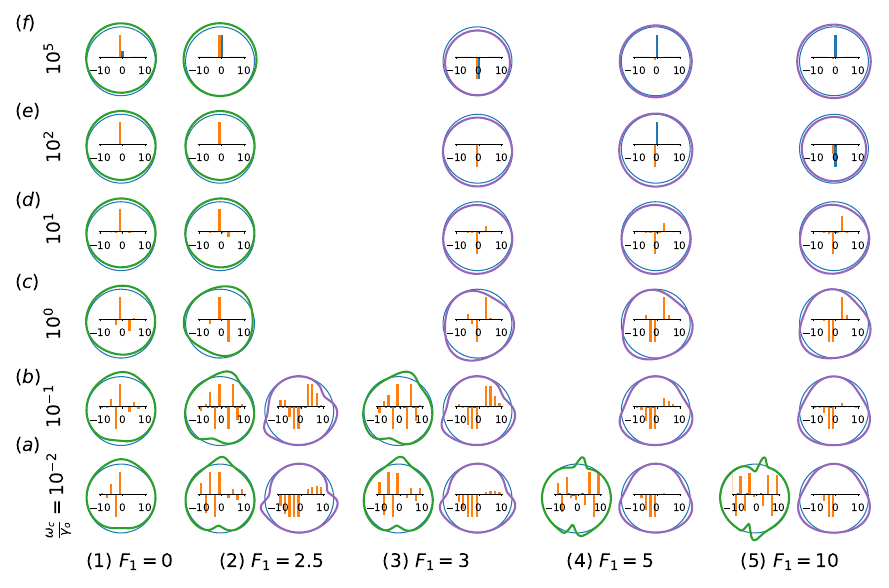}
    \caption{Fermi surface deformation \mbox{$h(\theta_{\bf p})$} for different Landau parameters (1--5) \mbox{$F_1= 0, \ 2.5, \ 3, \ 5, \ 10$} (left to right column) and magnetic fields (a)--(f) \mbox{$\omega_c/\gamma_o = 10^{-2}, \ 10^{-1}, \ 10^{0}, \ 10^{1}, \ 10^{2}, \ 10^{5}$} (bottom to top row). We use the same parameters as in Fig.~\ref{fig:3}, i.e.\ a constant odd-parity relaxation rate \mbox{$\gamma_o/\gamma_e =10^{-5}$} and set  \mbox{$v_Fk/\gamma_e = 5 \cdot 10^{-3}$} (\mbox{$k\xi = 1.58$}). We normalize the amplitude of the deformation to be approximately \mbox{$10\%$} of the Fermi surface. Green (purple) lines denote the upper (lower) branch deformation. The insets in each panel give the real-valued Fourier coefficients $a_m$ and $b_m$ [Eq.~\eqref{eq:expansion_coefficients}] in the \mbox{$m$}th angular momentum channel. We use the positive axis to denote cosine modes and the negative axis for the sine modes. We also indicate weight in the odd-parity (even-parity) sector by orange (blue) columns.}
    \label{fig:5}
\end{figure*}

\subsection{Structure of the Fermi surface}

To gain further insight on the disappearance of the tomographic collective modes, we examine the structure of the dimensionless Fermi surface deformation \mbox{$h(\theta_{\bf p})$} [defined in Eq.~\eqref{eq:h}] for the collective mode solutions. By the chain rule, this function describes a rigid angle-dependent shift of the chemical potential (and hence the Fermi surface) that will oscillate in space and time with the phase factor \mbox{$e^{-i\left(\omega t+{\bf k}\cdot {\bf r}\right)}$}. Figure~\ref{fig:5} shows the microscopic deformation of the upper and lower branch of the tomographic modes for various values of magnetic field [Fig.~\ref{fig:5}(a)--\ref{fig:5}(f) \mbox{$\omega_c/\gamma_o=10^{-2},\ 10^{-1}, \ 10^{0},\ 10^{1}, \ 10^{2}, \ 10^5$} and for different Landau parameters (1-5) \mbox{$F_1= 0, \ 2.5, \ 3, \ 5, \ 10$}. We otherwise use the same parameters as Fig.~\ref{fig:3}(a)--\ref{fig:3}(d) and normalize the amplitude of the deformation to  \mbox{$10\%$} of the Fermi energy. The insets in Fig.~\ref{fig:5} show histograms of the angular momentum decomposition of \mbox{$h(\theta_{\bf p})$} in terms of real-valued Fourier coefficients
\begin{equation}
    {\rm Re}\left[h(\theta_{\bf p})\right] = \sum_{m\geq 0}a_m \cos(m \theta) + \sum_{m>0}b_m \sin(m \theta) ,
    \label{eq:expansion_coefficients}
\end{equation}
where \mbox{$a_m = {\rm Re}[h_m+h_{-m}]$} and \mbox{$b_m = -{\rm Im}[h_m-h_{-m}]$}. We adopt a convention where the coefficients of \mbox{$a_m$} are plotted on the positive axis and \mbox{$b_m$} on the negative axis.

For very small magnetic fields (bottom row), the Fermi surface deformation associated with the upper branch (green lines) is peaked around \mbox{$\theta = \pm\pi/2$}. For \mbox{$F_1=0$} [Fig.~\ref{fig:5}(a1)], the peak is very broad, but becomes significantly narrower for finite~\mbox{$F_1$} [Fig.~\ref{fig:5}(a2)-\ref{fig:5}(a5)]. In this case, the increase in \mbox{$F_1$} leads to multiple sine  and cosine modes with odd \mbox{$m$} contributing to the upper branch collective mode. The structure of the lower branch (purple) in the absence of the magnetic field depends strongly on \mbox{$F_1$} as well. For \mbox{$F_c<F_1<F_1^*$} [Fig.~\ref{fig:5}(a2), \ref{fig:5}(a3)], the Fermi surface deformation is instead peaked around \mbox{$\theta \approx 0, \ \pi$} with an odd reflection symmetry around these points. The deformation associated with the lower branch involves not only a large number of odd-parity sine modes but also contains an admixture of odd-parity cosine modes in the low-magnetic field limit. As \mbox{$F_1$} increases past the critical value \mbox{$F_1^*$} [Fig.~\ref{fig:5}(a4), \ref{fig:5}(a5)], the Fermi surface deformation becomes broader, and more weight is placed in the sine modes with angular momentum  \mbox{$m= 1, \ 3$}. For sufficiently large \mbox{$F_1 \gg F_1^*$}, only the sine modes with \mbox{$m=1, \ 3$} contribute to the Fermi surface deformation at large~\mbox{$F_1$}.
With increasing magnetic field,
the deformations rotate and broaden. More weight is placed in the angular momentum modes \mbox{$m=\pm 1$}. At even larger magnetic fields [cf.~Fig.~\ref{fig:5}~(c-e)], the deformation of the Fermi surface primarily corresponds to the hydrodynamic current modes (\mbox{$m=\pm 1$}). This is expected as the limit of large magnetic fields can be shown to be equivalent to the hydrodynamic limit \mbox{$k\xi \ll 1$} \cite{Lee1975}, hence the mixing to higher angular momentum modes is suppressed at large fields.
Increasing the magnetic field further to \mbox{$\omega_c\sim \gamma_e$}, as in Fig.~\ref{fig:5}(f), ultimately leads to an increase in the hydrodynamic even-parity mode \mbox{$m=0$}. 

Figure~\ref{fig:5} highlights how the tomographic collective mode physics is destroyed in stages with increasing magnetic field. The first stage occurs when one of the branches of the tomographic modes disappears at finite magnetic field. However, this appears well within the tomographic regime, and the surviving tomographic collective mode are still composed of a large number of odd-parity Fermi surface modes. When \mbox{$\omega_c \sim \gamma_o$}, then the number of odd-parity modes contributing to the collective mode is reduced, ultimately leading to standard hydrodynamic transport and an admixture of even-parity Fermi surface modes for $\omega_c\sim \gamma_e$.

\section{Variational \textit{Ansatz} for the Tomographic Collective Mode}
\label{sec:variational}

The general features of the Fermi surface deformation in the absence of a magnetic field can be captured using a variational \textit{Ansatz}. To this end, we note that a formal solution to Eq.~\eqref{eq:kinetic_eqn_linearized} in terms of the dimensionless Fermi surface deformation \mbox{$h(\theta_{\bf p})$} [defined in Eq.~\eqref{eq:h}] has the form
\begin{equation}
    h(\theta_{\bf p}) = -v_Fe E_0 \hat{G} \sin(\theta_{\bf p}),
    \label{eq:formal_sol}
\end{equation}
where the inverse of the operator \mbox{$\hat{G}$} is given by
\begin{align}
    \hat{G}^{-1}h(\theta_{\bf p}) = &-i\omega h(\theta_{\bf p}) + i v_F k \cos(\theta_{\bf p}) \bar{h}(\theta_{\bf p}) + L[\bar{h}(\theta_{\bf p})]
\end{align}
with
\begin{equation}
    \bar{h}(\theta_{\bf p}) = h(\theta_{\bf p}) + 2F_1 \int_0^{2\pi} \frac{d\theta_{\bf p'}}{2\pi} \cos(\theta_{\bf p} - \theta_{\bf p'}) h(\theta_{\bf p'}).
\end{equation}
For general \mbox{$h(\theta_{\bf p})$} (i.e., for deformations that are not necessarily given by Eq.~\eqref{eq:formal_sol}), it is possible to show that the transverse conductivity tensor is bounded from below by~\cite{rostami_magnetic-field_2024}
\begin{equation}
    \sigma_T \geq \nu e^2 v_F^2(1+F_1)^2 \frac{|\langle \sin(\theta_{\bf p}) | h(\theta_{\bf p})\rangle|^2}{\langle h(\theta_{\bf p}) | \hat{G}^{-1} | h(\theta_{\bf p})\rangle},
    \label{eq:sigma_T_var}
\end{equation}
where we define the inner product
\begin{equation}
    \langle f(\theta_{\bf p}) | g(\theta_{\bf p}) \rangle = \int_0^{2\pi} \frac{d\theta_{\bf p}}{2\pi} f^*(\theta_{\bf p}) g(\theta_{\bf p}).
\end{equation}
The bound~\eqref{eq:sigma_T_var} becomes an equality if \mbox{$h(\theta_{\bf p})$} is the solution~\eqref{eq:formal_sol} to the linearized kinetic equation.

To proceed further, we split the Fermi surface deformation into odd- and even-parity modes,
\begin{equation}
    h(\theta_{\bf p}) = h_e(\theta_{\bf p}) + h_o(\theta_{\bf p}) ,
\end{equation}
where \mbox{$h_{e(o)}(\theta_{\bf p})$} denotes the even-parity (odd-parity) part of the Fermi surface deformation. In the absence of a magnetic field, there exists an exact relationship between the odd-parity and even-parity deformations~\cite{ledwith_tomographic_2019}
\begin{equation}
    h_e(\theta_{\bf p}) = \frac{-i v_Fk}{-i \omega + \gamma_e}\cos(\theta_{\bf p})\bar{h}_o(\theta_{\bf p}),
    \label{eq:even_ans}
\end{equation}
which implies that the odd-parity modes are suppressed by factors of \mbox{$\gamma_e$} in the tomographic limit. From this equation, the variational \textit{Ansatz} can be expressed solely in terms of the odd-parity deformation.

For the upper branch in the tomographic regime [see Fig.~\ref{fig:5}(a1)--\ref{fig:5}(3)], we use an odd-parity Gaussian \textit{Ansatz} with two peaks centered at \mbox{$\theta_{\bf p} = \pm \pi/2$} with a width \mbox{$\delta \theta$}
\begin{equation}
    h_o(\theta_{\bf p}) \sim e^{-\frac{(\theta_{\bf p}-\pi/2)^2}{\delta\theta^2}}- e^{-\frac{(\theta_{\bf p}-3\pi/2)^2}{\delta\theta^2}}.
    \label{eq:tom_ans_U}
\end{equation} 
For the lower branch, our numerical solution discussed above indicates that two separate trial ans\"atze are needed depending on the value of $F_1$. The first is for \mbox{$F_1 <F_1^*$} [Fig.~\ref{fig:5}(a2)]. In this regime, we use an odd-parity Gaussian \textit{Ansatz} centered at \mbox{$\theta_{\bf p} = 0, \pi$}:
\begin{equation}
    h_o(\theta_{\bf p}) \sim \theta_{\bf p}e^{-\frac{(\theta_{\bf p})^2}{\delta\theta^2}}+ (\pi-\theta_{\bf p}) e^{-\frac{(\theta_{\bf p}-\pi)^2}{\delta\theta^2}}.
    \label{eq:tom_ans_L_1}
\end{equation} 
As opposed to the upper branch, we also include an additional prefactor of \mbox{$\theta_{\bf p}$} to describe the asymmetry about the points \mbox{$\theta_{\bf p} = 0, \pi$}.
For larger values of \mbox{$F_1\gg F_1^*$} [Figs.~\ref{fig:5}(3)--\ref{fig:5}(5)], we observe that the deformation of the lower branch has equal weights of the \mbox{$m=\pm 1, \pm 3$} sine modes. For these values of \mbox{$F_1$}, it is more appropriate to use the \textit{Ansatz}
\begin{equation}
    h_o(\theta_{\bf p}) \sim \sin(\theta_{\bf p}) + \sin(3\theta_{\bf p}) 
    \label{eq:tom_ans_L_2}
\end{equation}
with no additional variational parameter.

The parameter \mbox{$\delta \theta$} in Eqs.~\eqref{eq:tom_ans_U}-\eqref{eq:tom_ans_L_1} is treated as a variational parameter that minimizes the transverse conductivity. Performing this procedure sets
\begin{equation}
    \delta \theta \approx \sqrt{\frac{2}{3}}\frac{1}{\sqrt{2 + \frac{(k\xi)^2}{(1-i \omega/\gamma_e)(1- i\omega/\gamma_o)}}} 
    \label{eq:delta_theta_U}
\end{equation}
for the upper branch, and (if \mbox{$F_1<F_1^*$})
\begin{equation}
    \delta \theta \approx \sqrt{\frac{12}{5}\frac{\omega^2 - (1+(k\xi)^2)\gamma_e \gamma_o +i \omega (\gamma_e + \gamma_o)}{2\omega^2 + (-2+(k\xi)^2) \gamma_e \gamma_o + 2 i \omega(\gamma_e + \gamma_o)}} 
    \label{eq:delta_theta_L}
\end{equation}
for the lower branch. Equations~\eqref{eq:delta_theta_U}--\eqref{eq:delta_theta_L} depend both on the frequency and the momentum of the collective mode. Note that in the zero-frequency limit at fixed \mbox{$k\xi$}, Eq.~\eqref{eq:delta_theta_U} is only a function of \mbox{$k\xi$} and vanishes as \mbox{$(k\xi)^{-1}$} for \mbox{$k\xi \gg 1$}, while in the limit  \mbox{$\omega/(v_Fk) \gg 1$} one finds \mbox{$\delta \theta \sim \omega/(v_Fk)$} becomes large and the \textit{Ansatz} of Eq.~\eqref{eq:tom_ans_U} is no longer applicable. We note that the zero-frequency limit of Eq.~\eqref{eq:tom_ans_U} and Eq.~\eqref{eq:delta_theta_U} can also be used to describe the scaling behavior of the static transverse conductivity in Ref.~\cite{rostami_magnetic-field_2024}, which implies that the deformation of the upper branch continuously extrapolates to the static limit. The structure of the lower branch is more complicated: in the zero frequency limit, Eq.~\eqref{eq:delta_theta_L} is imaginary for \mbox{$k\xi>2$}. This signals that this \textit{Ansatz} for the deformation is not appropriate for describing the conductivity at zero frequencies as it does not produce a real value. This is consistent with Fig.~\ref{fig:2}, which shows that the lower branch does not extrapolate to zero damping in the hydrodynamic limit, and instead disappears at a finite momentum.

\begin{figure}
    \centering
    \includegraphics{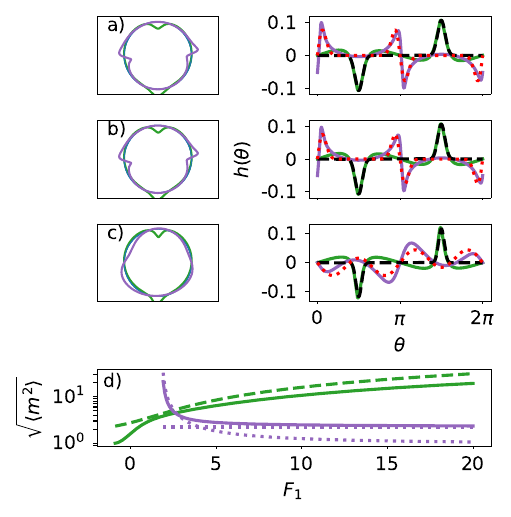}
    \caption{Fermi surface deformation of the upper and lower branch deformation at \mbox{$k\xi =1.58$} for (a)~\mbox{$F_1 = 2$}, (b)~\mbox{$F_1 = 2.6$}, and (c)~\mbox{$F_1 = 5$}. Left panels show polar plots of the Fermi surface deformation while right panels display the Fermi surface deformation as a function of the Fermi surface angle. In (a) and (b), black dashed and red dotted lines correspond to the variational \textit{Ansatz}, Eqs.~\eqref{eq:tom_ans_U} and \eqref{eq:tom_ans_L_1}, respectively. In (c) the lower branch \textit{Ansatz} is given by Eq.~\eqref{eq:tom_ans_L_2}. (d)~Root-mean-square of the angular momentum \mbox{$\sqrt{\langle m^2 \rangle}$} of the Fermi surface deformation of the collective mode [cf.\ Eq.~\eqref{eq:rms}] using the full numerical solution of the linearized kinetic theory (solid lines) and the variational \textit{Ansatz} (dashed lines), Eqs.~\eqref{eq:tom_ans_U}--\eqref{eq:tom_ans_L_2}.}
    \label{fig:6}
\end{figure}

To benchmark our variational results, in Fig.~\ref{fig:6}(a-c) we compare the variational \textit{Ansatz} with the numerically obtained expression for \mbox{$h(\theta_{\bf p})$} for different values of \mbox{$F_1$}, for the same parameters as in Fig.~\ref{fig:3}. The left column displays a polar plot of the Fermi surface deformation in the absence of a magnetic field for the both the upper (green) and lower (purple) branches of the tomographic mode obtained from the linearized kinetic theory. The right column shows the same deformation of the Fermi surface as a function of \mbox{$\theta$} alongside the results obtained from our variational \textit{Ansatz} for the upper branch (black dashed line) and lower branch (red dotted line). The width parameter \mbox{$\delta \theta$} is obtained using Eq.~\eqref{eq:delta_theta_U} (for the upper branch) and Eq.~\eqref{eq:delta_theta_L} (for the lower branch) at the exact value of the frequency of the collective mode  Eq.~\eqref{eq:delta_theta_U}. In general we find qualitatively good agreement between the exact deformation of the tomographic modes and the variational \textit{Ansatz} for all values of \mbox{$F_1$}.

For both Eqs.~\eqref{eq:tom_ans_U} and~\eqref{eq:tom_ans_L_1}, one can show that the root-mean-square of the angular momentum scales as \mbox{$\sqrt{\langle m^2 \rangle} \sim 1/\delta \theta$}, where the variance of the momentum is defined as
\begin{equation}
    \langle m^2 \rangle = \frac{\langle h(\theta_{\bf p}) | -\partial_{\theta_{\bf p}}^2 | h(\theta_{\bf p})\rangle}{\langle h(\theta_{\bf p})  | h(\theta_{\bf p})\rangle}.
    \label{eq:rms}
\end{equation}
For the lower-branch \textit{Ansatz} in Eq.~\eqref{eq:tom_ans_L_2}, we note that the variance is a constant \mbox{$\sqrt{\langle m^2 \rangle} = \sqrt{5}$}. In Fig.~\ref{fig:6}(d) we compare the root-mean-square of the angular momentum obtained from the numerically obtained Fermi surface deformation (solid lines) and the variational \textit{Ansatz} (dashed and dotted lines, respectively). For the upper branch the root-mean-square of the angular momentum obtained from the width in Eq.~\eqref{eq:tom_ans_U} qualitatively agrees with the numerical result for all values of~\mbox{$F_1$}. For small values of~\mbox{$F_1$}, the width of the lower branch agrees well with the Gaussian variational \textit{Ansatz} in Eq.~\eqref{eq:tom_ans_L_1}, while for large values of~\mbox{$F_1$}, the width saturates to a finite value in agreement with the prediction from Eq.~\eqref{eq:tom_ans_L_2}.

From these variational wavefunctions,
we can estimate the critical magnetic field observed in Fig.~\ref{fig:3}. To do so we note that the dominant odd-parity angular momentum modes in the microscopic deformation of the Fermi surface have angular momentum \mbox{$|m|\leq \sqrt{\langle m^2 \rangle}$}. When the odd-parity relaxation rate becomes smaller than the renormalized \mbox{$m$}th cyclotron resonance of one of these angular momentum modes (i.e.~\mbox{$\sqrt{\langle m^2 \rangle}\omega_c>\gamma_o(\sqrt{\langle m^2 \rangle})$}, the effect of the odd-parity relaxation becomes negligible compared to the cyclotron motion. Hence, if 
\begin{equation}
    \sqrt{\langle m^2 \rangle}\omega_c \sim \gamma_o(\sqrt{\langle m^2 \rangle}),
    \label{eq:critical_cond}
\end{equation}
the dominant modes of the Fermi surface deformation will be collisionless instead of tomographic.

We test this condition by plotting the critical magnetic field satisfying Eq.~\eqref{eq:critical_cond} for the upper (lower) branch as black dashed (dotted) lines in Fig.~\ref{fig:3} the estimates of the critical fields as obtained from the upper (dashed line) and lower branches (dotted line). There is good qualitative agreement between the disappearance of a tomographic mode and the critical magnetic field estimated from \mbox{$\sqrt{\langle m^2 \rangle}$} for both the staggered relaxation rate model [Eq.~\eqref{eq:constant_gammaodd}] and the \mbox{$m$}-dependent relaxation rates [Eq.~\eqref{eq:m_dependent_gammaodd}].

Increasing \mbox{$F_1$} leads to a larger variance for the upper branch, and thus the upper branch disappears at smaller magnetic fields. 
For \mbox{$F_c<F_1<F_1^*$} the variance of the angular momentum of the lower branch is larger than the upper branch which thus disappears earlier. At \mbox{$F_1= 2.6$} the upper and lower branches have equal widths. This point corresponds to  \mbox{$F_1^*$} where we observe both tomographic modes merge at a critical magnetic field, see Fig.~\ref{fig:3}(c).
When \mbox{$F_1>F_1^*$}, the upper branch has a larger \mbox{$\sqrt{\langle m^2\rangle}$} than the lower branch.  In this limit 
the upper branch disappears earlier with increasing magnetic field strength. Increasing \mbox{$F_1$} further, we see that the variance of the angular momentum of the upper branch continues to increase, while for the lower branch it saturates at \mbox{$\sqrt{\langle m^2\rangle}=\sqrt{5}$}.

In the limit of large \mbox{$k\xi$}, it is possible to determine \mbox{$F_1^*$} analytically by equating Eq.~\eqref{eq:delta_theta_U} and Eq.~\eqref{eq:delta_theta_L}
\begin{equation}
    \lim_{k \xi \to \infty} F_1^* = -1+\frac{2\sqrt{5}}{3(k\xi)^2} .
\end{equation}
Thus \mbox{$F_1^*\approx -1 + O\left((k\xi)^{-1}\right)$}. This implies that the upper branch will always disappear at a finite magnetic field while the lower branch persists for \mbox{$k\xi \gg 1$}. However, for \mbox{$k\xi\sim 1$}, as in Fig.~\ref{fig:3}, it is necessary to solve for~\mbox{$F_1^*$} numerically. For the parameters in Fig.~\ref{fig:3} this procedure gives \mbox{$F_1^*\approx 2.5$}, in good agreement with the result \mbox{$F_1^*\approx 2.6$} obtained above from solving the full linearized kinetic theory.

\section{Discussion}
\label{sec:Discussions}

In this article, we examined how diffusive collective modes in the tomographic regime are suppressed with the application of a magnetic field. In the absence of a magnetic field, the transverse conductivity exhibits two diffusive collective modes that are separated by a branch cut, the extent of which is approximately \mbox{$ -i \gamma_e (k\xi)^2 $} [cf.\ Eq.~\eqref{eq:frequencies_exact}]. Using our numerical solution of the linearized Boltzmann equation with even- and odd-parity relaxation rates, we explored how these tomographic collective modes change upon increasing the magnetic field for arbitrary Landau parameter \mbox{$F_1$}. The key finding is that, similar as for the static response~\cite{rostami_magnetic-field_2024, ben-shachar_tomographic_2025,ben-shachar_magnetotransport_2025}, there is a two-step suppression of the tomographic collective mode spectrum with increasing magnetic field. In the limit of small magnetic fields \mbox{$\omega_c \ll \gamma_o$}, there is no  appreciable change to the spectrum. The first stage occurs when the dominant odd-parity modes become collisionless, \mbox{$ \sqrt{\langle m^2\rangle}\omega_c \sim \gamma_o(\sqrt{\langle m^2\rangle})$}, at which point one branch of the tomographic modes disappears while the remaining mode persists. We note that at this stage the Fermi surface deformation of the remaining branch is still dominated by a large number of odd-parity angular momentum modes. The second stage occurs at larger magnetic fields \mbox{$\gamma_o \ll \omega_c \simeq \gamma_e$}, where the Fermi surface deformation becomes dominated by the hydrodynamic modes \mbox{$m=0,\pm1$}, and the damping rate approaches zero.

In principle, the tomographic modes could be observed in the transverse response of a material. Measuring shear sound and stress in condensed matter platforms is, however, difficult. Although there was a reported observation of shear sound in three-dimensional~\mbox{$^3$He}~\cite{Roach1976}, this has been debated~\cite{Flowers1976}. More recent proposals suggest to probe the current flow in narrow channels~\cite{Khoo2020}, or through enhancement of transmission of optical or terahertz radiation~\cite{Valentinis2021,Valentinis2021b}.~In principle, at finite magnetic fields, the mixing of longitudinal and transverse modes may be of benefit, since, tomographic and transverse sound modes will appear in standard longitudinal probes and the density response function, cf.\,~the discussion in Sec.~\ref{subsec:struct}. Although large magnetic fields will eventually suppress the tomographic regime, there may be an optimal parameter regime in which the longitudinal and transverse response are mixed, while not destroying the odd-even effect. However, our calculations indicate that the residue of the tomographic modes in the longitudinal response is negligible. 

Although our primary interest are applications in condensed matter systems, we note that our discussion applies equally well to atomic gases with synthetic gauge fields \cite{Dalibard2011,Aidelsburger2016}.
In particular, the odd-even effect also exist in two-dimensional fermionic atomic gases at low temperatures~\cite{maki24}, and our discussions can be readily carried over to these atomic gas platforms. Already there are experiments probing the longitudinal conductivity of a Fermi gas in an optical lattice by introducing a tilt to mimic a longitudinal electric field \cite{Anderson2019}. By oscillating the electric field in a direction perpendicular to the tilt, one could then extract the transverse conductivity response function, and search for signatures of the tomographic collective modes in the response.

\begin{acknowledgments}
We thank Denis Bandurin, Ulf Gran, Giovanni Vignale, and Oded Zilberberg for discussions. This work is supported by Vetenskapsr\aa det (Grant No.~2024-04485), the Olle Engkvist Foundation  (Grant No.~233-0339), the Knut and Alice Wallenberg Foundation (Grant No.~KAW 2024.0129), and Nordita.
\end{acknowledgments}

\appendix

\section{Evaluating \mbox{$x_{\pm 2}$}}\label{app:x_2}

\begin{figure}
    \centering
    \includegraphics{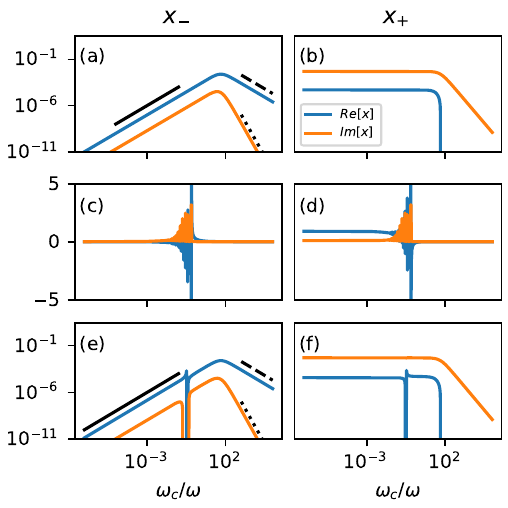}
    \caption{The functions $x_{\pm}(\omega, k)$ as a function of magnetic field. Here we have chosen \mbox{$v_F k/\omega=1$}. We use the staggered model for the relaxation rates [Eq.~\eqref{eq:constant_gammaodd}] for the linearized collision integral. The top, middle, and bottom rows correspond to (a)--(b) \mbox{$\gamma_e/\omega=\gamma_o/\omega = 10^2$}, (c)--(d) \mbox{$\gamma_e/\omega=\gamma_o/\omega = 10^{-2}$}, and (e)--(f) \mbox{$\gamma_e/\omega=10^4$}, \mbox{$\gamma_o/\omega = 10^{-2}$}, respectively. The black solid (dashed) lines in (a) and e) represent small (large) magnetic field asymptotics. For small magnetic fields $x_-$ vanishes as $\sim\omega_c$, while for large magnetic fields the real part vanishes as $\sim\omega_c^{-1}$, while the imaginary part as $\sim\omega_c^{-3}$.}
    \label{fig:7}
\end{figure}

In this appendix, we describe the evaluation of the function \mbox{$x_m$} defined in Eq.~\eqref{eq:continuedfraction}. For definiteness, we consider the case \mbox{$m>0$} (the extension to \mbox{$m<0$} is straightforward). Starting from Eq.~\eqref{eq:higher_m}, it is possible to obtain a recursive relation for \mbox{$x_m$}
\begin{align}
    &-i\left(\omega + i \gamma_m + m(1+F_m)\omega_c \right) \nonumber \\
    &+\frac{iv_Fk}{2}\left((1+F_{m+1}) x_{m+1} + (1 + F_{m-1}) \frac{1}{x_m}\right) = 0,
    \label{eq:app_higher_m}
\end{align}
which upon rearrangement gives
\begin{align}
    &x_m = \nonumber \\
    &\frac{-(1+F_{m-1} )\frac{iv_Fk}{2}}{-i (\omega + i \gamma_m + m(1+F_m)\omega_c ) + \frac{i v_Fk}{2}(1+F_{m+1})x_{m+1}}.
    \label{eq:app_x_m}
\end{align}

Eq.~\eqref{eq:app_x_m} is a continued-fraction representation of Eq.~\eqref{eq:app_higher_m}. Such continued fractions can be efficiently solved using Lentz's algorithm~\cite{Hofmann2022}. Once we have determined \mbox{$x_m$}, one can immediately infer the deviation of the Fermi surface in the \mbox{$m$}th angular-momentum sector [Eq.~\eqref{eq:expansion_n}]
\begin{equation}
    h_m = x_m x_{m-1} \cdot ... \cdot x_2 \, h_1.
    \label{eq:app_deformations}
\end{equation}

Consider now \mbox{$x_{\pm2}$}, which enter directly into the reduced kinetic theory for the hydrodynamic modes in Eq.~\eqref{eq:reduced_kinetic_theory}. In the hydrodynamic limit, \mbox{$v_Fk\ll\gamma_e,\gamma_o$}, it is possible to truncate the continued fraction for \mbox{$x_{\pm2}$} to lowest order in \mbox{$vFk/\gamma_e$}. This is equivalent to only retaining the \mbox{$m=0,\pm1, \pm 2$} modes in the linearized kinetic theory in Eq.~\eqref{eq:Boltzmann_m_E}. Defining \mbox{$x_{\pm} = x_2 \pm x_{-2}$}, one obtains at this level of approximation
\begin{align}
    x_+ &= v_Fk \frac{(1+F_1)(\omega+ i \gamma_e)}{(\omega + i\gamma_e)^2-  4\omega_c^2} \label{eq:xplus_smallfield}, \\[0.5ex]
    x_- &= v_F k \frac{2(1+F_1)\omega_c}{(\omega + i\gamma_e)^2-  4\omega_c^2}.
    \label{eq:xminus_smallfield}
\end{align}
As discussed in the main text, in the limit of small magnetic fields \mbox{$x_-$} vanishes linearly with \mbox{$\omega_c$} while \mbox{$x_+$} remains finite.
Interestingly, Eqs.~\eqref{eq:xplus_smallfield}--\eqref{eq:xminus_smallfield} have the same structure as the finite-frequency shear and Hall viscosity, $\eta_S(\omega)$ and $\eta_H(\omega)$, calculated under the approximation that the hydrodynamic modes only couple to the \mbox{$m=2$} modes~\cite{afanasiev_shear_2023}. In fact, a direct comparison with Ref.~\cite{afanasiev_shear_2023} gives
\begin{align}
\eta_s(\omega) &=\lim_{k \to 0} i \frac{v_F}{4k} x_+, \\
\eta_H(\omega) &= \lim_{k \to 0} \frac{v_F}{4k} x_-.
\label{eq:viscosity_x_relation}
\end{align}
In other words, the \mbox{$x_{\pm}$} are directly proportional to the \mbox{$k=0$} shear and Hall viscosity spectral functions \cite{Alekseev2018, afanasiev_shear_2023, khoo_shear_2019}.

In the limit of large magnetic fields with finite momentum and frequency, the continued fraction has the form:
\begin{align}
    x_2 \approx \frac{-(1+F_{1}) \frac{iv_Fk}{2}}{-i 2\omega_c + \frac{\frac{i v_Fk}{2}}{-i(1+F_3)3\omega_c +... }},
\end{align}
and similarly for \mbox{$x_{-2}$}. If all Landau parameters \mbox{$F_{m\geq 1}$} are zero, an exact solution for~\mbox{$x_{\pm2}$} can be obtained, which reads
\begin{equation}
    x_{\pm2} \approx \mp \frac{z(1+F_1)}{2}\frac{_0F_1(3;-z^2)}{_0F_1(2;-z^2)},
    \label{eq:app:x_2_large_B}
\end{equation}
where \mbox{$z=v_Fk/(2\omega_c)$} and \mbox{$_aF_b(c;z)$} is  a hypergeometric function. Evaluating Eq.~\eqref{eq:app:x_2_large_B} to first order in \mbox{$1/\omega_c$} gives 
\begin{align}
    x_+ & \approx 0 \\[0.5ex]
    x_- &\approx -\frac{(1+F_1)v_Fk}{2\omega_c}.
    \label{eq:asymptotic_xpm}
\end{align}
This result can also be obtained from Eq.~\eqref{eq:xplus_smallfield}--\eqref{eq:xminus_smallfield} after subsequently taking the limit of large cyclotron frequencies. Thus the limit of large magnetic fields is equivalent to the hydrodynamic limit. This was first noted in Refs.~\cite{Lee1975,Simon1993} by solving the Fermi-liquid kinetic theory in a frame rotating at the quasiparticle's cyclotron frequency.

For finite magnetic fields, it is necessary to evaluate the continued fraction numerically. The numerical solution to \mbox{$x_{\pm}$} is shown in Fig.~\ref{fig:7} for the relaxation rates given in Eq.~\eqref{eq:relaxation_rates} with a constant odd-parity relaxation rate, Eq.~\eqref{eq:constant_gammaodd}. In the small magnetic-field limit, $x_+$ approaches a constant while \mbox{$x_-$} vanishes as ${\it O}(\omega_c)$. In the limit of large magnetic fields, the real and imaginary parts of \mbox{$x_{-}$} decay as \mbox{$\omega_c^{-1}$} and \mbox{$\omega_c^{-3}$}, respectively. Likewise, for \mbox{$x_+$}, the real part approaches a constant, and the imaginary part decays as a power law that scales as \mbox{$\omega_c^{-2}$}. 

Depending on the values of \mbox{$\gamma_m$}, we observe different qualitative features in \mbox{$x_{\pm}$}: for \mbox{$\gamma_m \gg \omega$} [Fig.~\ref{fig:7} (a)-(b)], \mbox{$x_{\pm}$} smoothly crosses over from a small magnetic-field  to a large-field regime. There are no branch cuts or discontinuities appearing that would signal the presence of the tomographic modes. This is indicative of the conventional hydrodynamic and collisionless regimes. In the opposite limit [Fig.~\ref{fig:7} (c)-(d)], there are substantial oscillations in \mbox{$x_{\pm}$} whenever \mbox{$\omega\approx \omega_c$}. Finally in Fig.~\ref{fig:7}(e-f), there is an odd-even effect. When the frequency satisfies \mbox{$\gamma_e \gg \omega$} while \mbox{$\gamma_o \ll \omega$}, a branch cut emerges in $x_{\pm}$. This branch cut is responsible for the two tomographic diffusive poles in the transverse conductivity in the zero magnetic field limit~\cite{Hofmann2022}.

\bibliography{Bibliography} 

\end{document}